\pdfoutput=1
\documentclass[fleqn,usenatbib]{mnras}

\usepackage{newtxtext,newtxmath}

\usepackage[T1]{fontenc}
\usepackage{ae,aecompl}

\usepackage{graphicx}	
\usepackage{amsmath}	
\usepackage{amssymb}	

\title[$^{7}$Be in the nova ASASSN-16kt]{Beryllium detection in the very fast nova ASASSN-16kt (V407 Lupi)}

\author[L. Izzo et al.]{
L. Izzo,$^{1}$\thanks{E-mail: izzo@iaa.es}
P. Molaro,$^{2}$
P. Bonifacio,$^{3}$
M. Della Valle,$^{4,5}$\newauthor
Z. Cano,$^{1}$
A. de Ugarte Postigo,$^{1,6}$
J. L. Prieto,$^{7,8}$
C. Th\"one,$^{1}$\newauthor
L. Vanzi,$^{9}$ A. Zapata,$^{9}$ D. Fernandez,$^{10}$
\\
$^{1}$Instituto de Astrofisica de Andalucia (IAA-CSIC), Glorieta de la Astronomia s/n, E-18008 Granada, Spain \\
$^{2}$INAF-Osservatorio Astronomico di Trieste, Via G.B. Tiepolo 11, I-34143 Trieste, Italy\\
$^{3}$GEPI, Observatoire de Paris,  PSL Research University, CNRS, Place Jules Janssen, 92195 Meudon, France\\
$^{4}$INAF-Osservatorio Astronomico di Napoli, Salita Moiariello, 16, I-80131 Napoli, Italy\\
$^{5}$International Center for Relativistic Astrophysics, Piazza della Repubblica 10, I-65122 Pescara, Italy\\
$^{6}$Dark Cosmology Center, Niels-Bohr-Institute, University of Copenhagen, Juliane Maries Vej 30, DK-2100 Copenhagen, Denmark\\
$^{7}$N\'ucleo de Astronom\'ia de la Facultad de Ingenier\'ia y Ciencias, Universidad Diego Portales, Av. Ej\'ercito 441, Santiago, Chile\\
$^{8}$Millennium Institute of Astrophysics, Santiago, Chile\\
$^{9}$Department of Electrical Engineering and Center of Astro Engineering, Pontificia Universidad Catolica de Chile\\ Av. Vicuna Mackenna 4860 Santiago, Chile\\
$^{10}$Institute of Astrophysics, Pontificia Universidad Catolica de Chile, Av. Vicuna Mackenna 4860 Santiago, Chile\\
}

\date{Accepted XXX. Received YYY; in original form ZZZ}

\pubyear{2017}

\begin{document}
\label{firstpage}
\pagerange{\pageref{firstpage}--\pageref{lastpage}}
\maketitle

\begin{abstract}
We present high-resolution spectroscopic observations of the fast nova ASASSN-16kt (V407 Lup). A close inspection of spectra obtained at early stages has revealed the presence of low-ionization lines, and among the others we have identified the presence of the ionised $^7$Be doublet in a region relatively free from possible contaminants. After studying their intensities, we have inferred that ASASSN-16kt has produced (5.9 - 7.7)$ \times 10^{-9}$ M$_{\odot}$ of $^7$Be. The identification of bright Ne lines may suggest that the nova progenitor is a massive (1.2 M$_{\odot}$) oxygen-neon white dwarf. The high outburst frequency of oxygen-neon novae implies that they likely produce an amount of Be similar, if not larger, to that produced by carbon-oxygen novae, then confirming that classical novae are among the main factories of lithium in the Galaxy.
\end{abstract}

\begin{keywords}
nuclear reactions, nucleosynthesis, abundances -- stars: individual: ASASSN-16kt (V407 Lup) -- novae, cataclysmic variables -- Galaxy: evolution
\end{keywords}



\section{Introduction}

During a nova outburst, the thermo-nuclear runaway (TNR) process produces several beta isotopes of neon, oxygen and fluorine whose decays release enough energy to eject the accreted layers \citep{Wagoner1967,Fowler1975,Starrfield1996,Hernanz1996,Starrfield1998,Jose1998,Yaron2005,Starrfield2009}. At the same time, they furnish the nova ejecta with non-solar CNO isotopes (mainly $^{13}$C, $^{15}$N and $^{17}$O) that will propagate into the interstellar medium then contributing to the Galactic chemical enrichment \citep{Gehrz1998,Jose2004}. It was proposed by \citet{CameronFowler1971} that during advanced phases of red giant stars the reaction of $^3$He with $^4$He can give rise to the $^7$Be isotope, which decays only through electron capture into Li after a half-time decay of $\sim$ 53 days \citep{Giraud2007}. This mechanism was later applied to novae by \citet{Arnould1975}. Consequently, classical novae have been proposed as one of the possible lithium factories in the Galaxy \citep{Starrfield1978,DAntonaMatteucci1991,Romano1999}.

In recent years, thanks to the use of high-resolution spectrographs, we were for the first time able to observe the presence of $^7$Li in the early spectra of novae V1369 Cen \citep{Izzo2015} and V5668 Sgr (Wagner et al. in preparation, Izzo et al. in preparation). In addition, evidence of large quantities of $^7$\ion{Be}{II} has also been reported, after $\sim$ 60 days from the outburst, for the carbon-oxygen (CO) novae V339 Del \citep{Tajitsu2015}, V2944 Oph \citep{Tajitsu2016}  and V5668 Sgr \citet{Molaro2016}. In the latter case, the observation of unsaturated narrow components allowed an accurate  $^7$\ion{Be} abundance determination, which led the authors to suggest that classical novae are one of the main lithium factories in the Galaxy.  

In this paper we present a new detection of $^7$\ion{Be}{II} in the spectra of ASASSN-16kt (also named V407 Lup), an oxygen-neon (ONe) nova, then providing additional support that ONe novae also produce $^7$\ion{Be}{II}. After presenting our dataset (Section 2), we discuss some physical properties of the nova system such as its distance and the extinction, as inferred from the spectral dataset (Section 3). Then, we present an analysis of the early phases of the nova, showing the presence of $^7$\ion{Be}{II} and the absence of $^7$\ion{Li}{I} in the early spectra (Section 4). Finally, from the unique late nebular spectrum in our dataset, we provide an estimate of the mass of the nova ejecta and hence the amount of $^7$\ion{Be}{II} produced in the nova outburst (Sections 5 and 6), before drawing our conclusions in Section 7.

\section{Observations}

\begin{figure}
	\includegraphics[width=1.0\columnwidth]{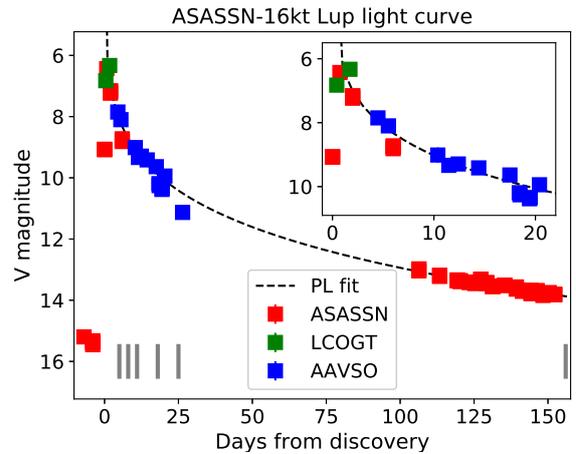}
    \caption{$V$ light curve of ASASSN-16kt obtained with data from ASAS-SN, LCOGT and AAVSO observations. The dashed black line represents the best-fit of the data (excluding the discovery observation) with a power-law function, used to estimate the $t_2$ value. The gray vertical lines correspond to the spectroscopical observation epochs. The gap of data between days 40 and 100 is mainly due to the difficult visibility of the nova when it was too close to the Sun. The inset figure shows a zoom around the first 20 days of the nova outburst.}
    \label{fig:1}
\end{figure}

ASASSN-16kt was discovered by the All-Sky Automated Survey for SuperNovae \citep[ASAS-SN)][]{Shappee2014} as a bright source of $V=9.1$ mag on Sep 24 2016 \citep{ATEL9538}. The nova brightened in the following two days, reaching a maximum of $V = 6.3$ on Sep 26 2016 \citep{ATEL9550}, after which it started to decay rapidly. In Fig.\ref{fig:1} we show the $V$ mag light curve of ASASSN-16kt obtained with data from the ASAS-SN \citep{Kochanek2017}, LCOGT \citep{Brown2013} and AAVSO\footnote{http://www.aavso.org} observations. If we exclude the first data points, corresponding to the pre-discovery and when the nova was still rising in luminosity, the light curve (magnitude vs time) is well fitted with a simple power-law function with power-law index $\gamma = -0.16 \pm 0.01$. From this we derive a value for the time the light curve decays of two and three magnitudes from the peak of $t_2 = 9.9 \pm 0.1$ days and $t_3 = 19.4 \pm 0.1$ days.

After the discovery, we started a DDT program at ESO-VLT with UVES and X-Shooter (Program ID 297.D-5065(B), PI L. Izzo). In parallel, we observed the nova with the PUC High Echelle Resolution Optical Spectrograph \citep[PUCHEROS,]{Vanzi2012} mounted on the ESO 0.5~m telescope located at the Observatory of Pontificia Universidad Catolica (OUC) in Santiago, Chile. We obtained a total of six epochs, five of these concentrated in the first 25 days and the last one after 155 days from its discovery, see also Fig. \ref{fig:1}. The detailed log of the observations is shown in Table \ref{tab:1}. PUCHEROS data have been reduced with the CERES routine for echelle spectra \citep{CERES}. The ESO-VLT UVES data have been reduced and flux-calibrated using the \textit{reflex} environment \citep{reflex}, while for the X-shooter data we used the standard ESO Recipe Execution Tool (\textit{esorex})\footnote{http://www.eso.org/sci/software/cpl/esorex.html}. We made extensive use of various python scripts for the analysis and IRAF packages for counter-checking all measurements. In particular we recognize the use of the \texttt{numpy} \citep{Numpy}, \texttt{matplotlib} \citep{Matplotlib} and the \texttt{astropy} \citep{Astropy2013} packages. 

\begin{table}
\centering
\caption{Log of observations. The values on the first column refer to days from the discovery.}
\label{tab:1}
\begin{tabular}{l c c c c}     
\hline
Epoch & Date & Instrument & Range (\AA)\\
\hline
Day 5 & Sep. 29 2016 & PUCHEROS & 4250 - 6900\\
Day 8 & Oct. 2 2016 & UVES & 3060 - 9460\\
Day 11 & Oct. 5 2016 & PUCHEROS & 4250 - 6900\\
Day 18 & Oct. 12 2016 & UVES & 3060 - 9460\\
Day 25 & Oct. 19 2016 & X-Shooter & 3140 - 24800\\
Day 155 & Feb. 27 2017 & X-Shooter & 3140 - 24800\\
\hline
\end{tabular}
\end{table}

\section{General Properties}

According to the classification introduced by \citet{Payne1957}, ASASSN-16kt can be considered a very fast nova ($t_2 \leq 10$ days). CNe undergo a super-Eddington phase \citep[see][ and references therein]{DellaValle1995} during their maximum luminosity. This is a parameter which depends on the temperature of the progenitor WD, $T_{WD}$, on the accretion rate, $\dot{M}$ \citep{Yaron2005}, but mainly on the mass of the WD progenitor \citep{Livio1992}: bright and fast novae are indeed generally related to more massive progenitors \citep{IbenTutukov1985}. Moreover, a massive WD needs less accreted matter to initiate the TNR, and consequently the mass ejected in the outburst is less that expelled by less massive nova progenitors. An immediate result of this theoretical deduction is the existence of some correlations between the absolute magnitude at nova maximum and the rate of decline \citep[the Maximum-Magnitude Rate-of-Decline, MMRD, see also][]{Downes2000}. However, we are well aware that the accuracy provided by the MMRD methods used for novae in M31 and LMC is of the order of 30$\%$ \citep{VanDenBergh1992,DellaValle1995}, and that not all novae follow these correlations, like the new class of very fast and faint novae with a $t_2 \sim$ of few hours for which an extension of the MMRD was pointed out \citep{Kasliwal2011}. In what follows, we take into account this additional source of uncertainty.

Using the MMRD formulation given in \citet{DellaValle1995}:
\begin{equation}
V_{pk} = -7.92 - 0.81 \times \arctan \Big(\frac{1.32 - \log t_2}{0.23}\Big),
\end{equation}
with $V_{pk}$ the absolute V magnitude at nova maximum and $t_2$ the time the nova decay of 2 mag from the maximum, we find the distance of ASASSN-16kt to be $d = 10.0 \pm 1.5$ kpc. Similar results are obtained using the \citet{Buscombe1955} method, in the formulation given by \citet{Downes2000}, where all novae at 15 days after the maximum have a similar absolute magnitude, given by $M_{V,15} = -6.05 \pm 0.44$. For ASASSN-16kt we have that the observed $V$-band magnitude at 15 days is $V_{15} = 9.6$ mag, then we obtain a distance of $d_{V,15} = 11.2 \pm 2.0$ kpc, in agreement with our previous estimate.

At galactic coordinates of $l = 330.0938$\textdegree and $b = 9.573$\textdegree, we obtain a height above the galactic plane of $h = 1700 \pm 300$ pc, which is typical for \textit{bulge} novae \citep{DellaValle1998}. We also report that the position of the nova is 0.5$''$ distant from a GAIA DR1 source (ID 5999691733347769472) with a still unknown parallax and a magnitude of $G = 18.95$. Assuming this source is the possible progenitor of ASASSN-16kt, we estimate an amplitude of the outburst of $A \sim 12.3 \pm 0.2$. We finally note that, given the above estimate of the distance, the lack of a measured parallax in the DR1 is consistent with our result for the distance of the nova. 

At maximum light, novae show an intrinsic color index of $\langle B - V \rangle_{\textrm{max}} = 0.23 \pm 0.06$,  \citep{VanDenBergh1987}. This is explained by the reprocessing of the incoming radation from the underlying WD at longer wavelengths through the nova ejecta \citep{ShoreBASI}. Consequently, we can estimate the intrinsic extinction from the observed color index: at maximum, ASASSN-16kt has $V = 6.33 \pm 0.06$ and $B = 6.86 \pm 0.06$ \citep{ATEL9550}, and hence $(B - V)_{\textrm{obs}} = 0.53 \pm 0.08$. Combining this observed data with the results of \citet{VanDenBergh1987} we get an intrinsic extinction of $E(B-V) = 0.30 \pm 0.09$.

\begin{figure}
	\includegraphics[width=1.0\columnwidth]{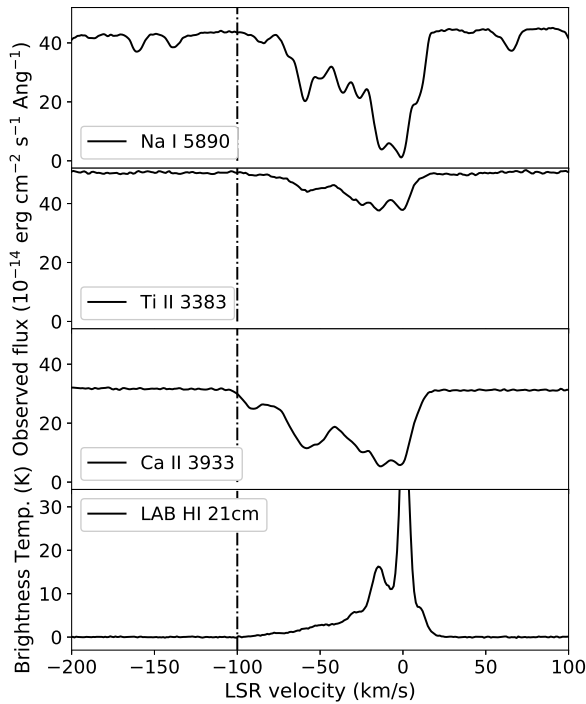}
    \caption{The structured line profiles of the IS lines \ion{Na}{I} 5890, \ion{Ti}{II} 3383 and \ion{Ca}{II} 3933 compared with the profile of the \ion{H}{I} 21 cm line from the LAB survey \citep{Kalberla2005}, all corrected for the local standard of rest (LSR) velocity. The black dashed line marks the $v = -100$ km s$^{-1}$ limit, corresponding to the beginning of the rising line profile, and implying that the nova is located at the border of the galactic disk.}
    \label{fig:2}
\end{figure}

The high resolution provided by UVES allowed us to identify several interstellar (IS) lines and diffuse interstellar bands (DIBs) in the first UVES spectrum (Day 8). Several IS lines such as \ion{Na}{I} D, \ion{Ti}{II} and \ion{Ca}{II} H,K, have similar profiles with multiple absorption features at bluer wavelengths, see Fig. \ref{fig:2}. The common maximum blue-shifted velocity reported for DIBs is $v \sim -100$ km s$^{-1}$. These observations point to the presence of a number of IS clouds along the line-of-sight to the nova, and the observed maximum blue-shifted components suggests them to be at a large distance from us. In Fig. \ref{fig:2} we also show the neutral hydrogen 21 cm line profile corrected to heliocentric velocities, as observed by the LAB survey \citep{Kalberla2005}, at the position of the nova. We note an increasing brightness temperature in the \ion{H}{I} 21 cm profile starting exactly at $v_{\textrm{hel}} \eqsim -100$ km s$^{-1}$, suggesting that, considering its galactic coordinates, the nova is located at the opposite border of the Galactic disk, with respect to the Sun location. 

Some DIBs can also be used to estimate line-of-sight reddening. To this aim, we used the correlations between the equivalent widths (EW) of these DIBs, the column density of neutral atomic hydrogen $N(H)$ and the color excess provided by \citet{Friedman2011}. The total list of IS lines and DIBs with secure identification and measurements of their observed central wavelengths as well as of their EWs is reported in Table \ref{tab:2}. We obtained an estimate for  N(H) and E(B-V) using DIBs at $\lambda\lambda$ 5780, 5797, 6196 and 6614. After a first check, we excluded DIB $\lambda$ 6196 since the estimated N(H) value is larger that derived from an analysis of the neutral hydrogen line at 21 cm from the LAB survey, i.e. $N(H)_{\textrm{HI}} = 1.54 \times 10^{21}$ cm$^{-2}$: we believe indeed that this DIB is blended with some other unidentified component. The final values obtained from the average of the remaining DIB estimates are: $N(H) = 1.03 \times 10^{21}$ cm$^{-2}$ and $E(B-V) = 0.24 \pm 0.02$, which is in good agreement with our above estimate obtained from the observed color index at maximum. We then used these values to correct the spectra for Galactic extinction.

\begin{table}
\centering
\caption{List of observed IS lines and DIBs in the Day 8 UVES spectrum. We give the IS/DIB name (col. 1), the observed central wavelength (2), the observed EW (3), and the rest-frame DIB wavelength (4). For the \ion{Na}{I} 3302/3 doublet, we report the total EW of both components.}
\label{tab:2}
\begin{tabular}{l c c c}     
\hline
DIB & $\lambda_{\textrm{obs}}$ & EW & $\lambda_{\textrm{rf}}$ \\
 & (\AA) & (m\AA) & (\AA) \\
\hline
\ion{Ti}{II} 3242 & 3242.22 & 101.0$\pm$5.8 & 3241.994  \\
\ion{Na}{I} 3302/3 & 3302.60/3.21& 61.2$\pm$3.8 & 3302.369/.978  \\
\ion{Ti}{II} 3383 & 3384.00 & 152.1$\pm$6.2 & 3383.768  \\
\ion{Ca}{II} 3933 & 3933.91 & 718.5$\pm$43.7 & 3933.663  \\
\ion{Ca}{II} 3968 & 3968.72 & 487.1$\pm$14.5 & 3968.469  \\
\ion{Na}{I} 5890 & 5890.33 & 881.6$\pm$16.9 & 5889.950  \\
\ion{Na}{I} 5896 & 5896.31 & 695.2$\pm$10.9 & 5895.924  \\
\hline
CH+ 3958 & 3957.97 & 47.3$\pm$1.0 & 3957.7  \\
CH+ 4232 & 4232.85 & 13.7$\pm$0.4 & 4232.6  \\
CH 4300 & 4300.65 & 7.8$\pm$0.2 & 4300.3  \\
DIB 5780 & 5780.69 & 105.5$\pm$0.9 & 5780.5  \\
DIB 5797 & 5797.36 & 50.5$\pm$0.8 & 5797.1  \\
DIB 6196 & 6196.39 & 25.1$\pm$0.4 & 6196.0  \\
DIB 6614 & 6613.97 & 50.8$\pm$0.3 & 6613.6  \\
\hline
\end{tabular}
\end{table}

A summary of the main properties of ASASSN-16kt is shown in Table \ref{tab:summary}

\begin{table}
\centering
\caption{Summary of the main physical properties of ASASSN-16kt: 1) the V magnitude at maximum brightness $V_{max}$; 2) the amplitude $A$ in magnitudes from the possible progenitor star; 3-4) the $t_2$ and $t_3$ values; 5) the color index $E(B-V)$; 6) the maximum ejecta velocity $v_{ej}$; 7) the distance $d$ and 8) the height $h$ above the galactic plane.}
\label{tab:summary}
\begin{tabular}{l r }     
\hline
Property & Value \\
\hline
$V_{max}$ & $6.3$ mag\\
$A$ & 12.3 $\pm$ 0.2 mag\\
$t_2$ & 9.9 $\pm$ 0.1 days\\
$t_3$ & 19.4 $\pm$ 0.1 days\\
$E(B-V)$ & 0.24 $\pm$ 0.02 mag\\
$v_{ej}$ & 2370 km s$^{-1}$\\
$d$ & 10.0 $\pm$ 1.5 kpc\\
$h$ & 1.7 $\pm$ 0.3 kpc\\
\hline
\end{tabular}
\end{table}

\section{The early spectra}

\begin{figure*}
	\includegraphics[width=18cm]{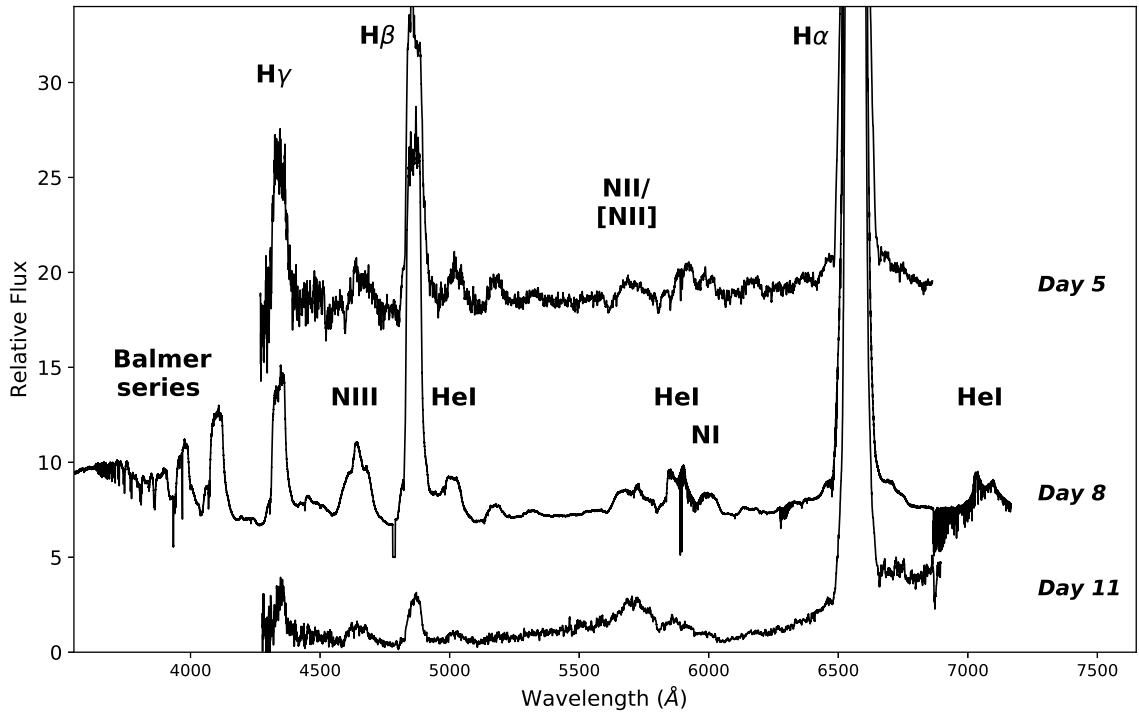}
    \caption{The optical spectra of ASASSN-16kt observed on Day 5 and Day 11 with PUCHEROS and on Day 8 with UVES. We show only a limited wavelength range (3550-7300 \AA) of the UVES spectrum for clarity. He/N features are reported and visible in both spectra where coverage with PUCHEROS is available. A faint emission line attributed to \ion{Fe}{II} $\lambda$5168 is clearly visible in the first two (Day 5 and Day 8) spectra.}
    \label{fig:3}
\end{figure*}

The first spectroscopic observation of the nova was obtained with PUCHEROS ($R \sim 20000$) five days after the nova discovery, see Fig. \ref{fig:3}. The spectrum shows clear Balmer series P-Cygni profiles with broad blue-shifted absorptions centered at $v = -2030,-3650$ km s$^{-1}$. The first of these absorption systems coincides in velocity with the observed \ion{Na}{I} D P-Cygni profile line. The presence of \ion{N}{III} $\lambda$4640 and \ion{He}{I} $\lambda$5016 in emission suggests a He/N spectral class for the nova \citep{Williams1991a}. However, the presence of an emission line coincident with \ion{Fe}{II} $\lambda$5169 tells us that the nova could have passed through a very fast iron-curtain phase. 

The following spectrum, obtained three days later with UVES at VLT, confirms this hypothesis. We clearly distinguish emission lines of \ion{He}{I} $\lambda\lambda$ 5016,5876, \ion{N}{II} $\lambda$5678 and the blend \ion{N}{III} $\lambda$4640 as the strongest non-Balmer lines, if we exclude the \ion{O}{I} $\lambda$8446, see Fig. \ref{fig:3}. We also identify transitions of \ion{N}{I} quartet lines $\lambda\lambda$7454, 8212, 8692, and \ion{N}{I} doublet lines $\lambda\lambda$7904,9028/60,9395 (partly embedded with telluric lines), as well as \ion{N}{I} $\lambda$6005. All these neutral nitrogen lines are characterised by the absence of P-Cygni profiles and a bell-shaped emission-line profile, unlike the \ion{He}{I} lines which show a clear saddle-shaped profile lines, with evidence of P-Cygni absorption at $v_{ej}$ = -2030 and -3830 km s$^{-1}$. We also note the presence of forbidden lines like [\ion{N}{II}] $\lambda$5755, which is partially blended with the \ion{N}{II} 5678 line, and faint [\ion{O}{II}] $\lambda\lambda$7320/30 doublet, which is embedded in telluric absorption lines. These features are typical of the He/N spectral class.

\begin{figure}
	\includegraphics[width=1.0\columnwidth]{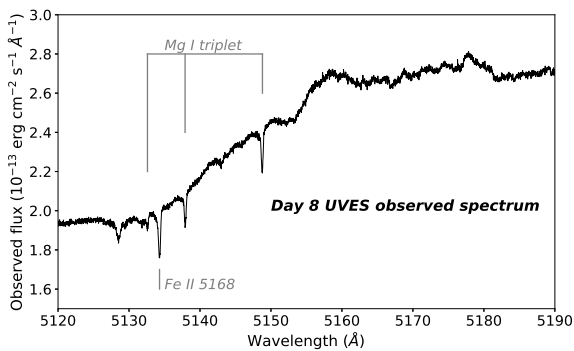}
    \caption{Day 8 observed spectrum in the range 5120-5190 \AA. The \ion{Fe}{II} $\lambda$5168 line and the \ion{Mg}{I} triplet coincide with their laboratory wavelengths, after correcting for the blue-shifted ejecta velocity as measured from \ion{Na}{I} D lines. The additional absorption observed at $\lambda$5128.5 corresponds to the blue-shifted component at $v_{\rm{exp,2}} = -2370$ km s$^{-1}$ of \ion{Fe}{II} $\lambda$5168, see also Fig. \ref{fig:5}.}
    \label{fig:4}
\end{figure}

The high resolution, high signal-to-noise data provided by UVES also allow us to identify low-ionization elements in the nova ejecta. The \ion{Na}{I} D lines serve as reference for the identification of low-ionization transitions \citep[see e.g. the method developed by ][]{Williams2008}: we determine an expansion velocity from analysis of \ion{Na}{I} D lines of $v_{ej,1} = -2030$ km s$^{-1}$, similar to what we have observed for \ion{He}{I} $\lambda$5876, and an additional fainter component at $v_{ej,2} = -2370$ km s$^{-1}$. Assuming the first value as the ejecta velocity of the low-ionization elements, we are able to identify two lines of \ion{Fe}{II} (multiplet 42) $\lambda\lambda$ 5018,5169  (the P-Cygni of the other line at 4924 \AA\, is not clearly identified, since the corresponding absorption is  located within the red wing of the bright H$\beta$ line) and the \ion{Mg}{I} $\lambda$5178 triplet, see Fig. \ref{fig:4}. Furthermore, we clearly identify \ion{Fe}{II} $\lambda$5018 and other similar transitions at $\lambda\lambda$3154,3168,3210,3213,3228, the triplet of \ion{Mg}{I} $\lambda\lambda$3829,3832,3838, \ion{Ca}{II} $\lambda\lambda$3933,3968 and several lines of \ion{Cr}{II} $\lambda\lambda$3132,3135,3137,3147. Concluding, ASASSN-16kt shows traces of low-ionization elements that are typical of the \ion{Fe}{II} nova spectral class (according to the Cerro-Tololo classification) and confirms the end of a very rapid iron-curtain phase.

\subsection{Detection of the $^7$Be II $\lambda$3130 doublet}

\begin{figure*}
	\includegraphics[width=18cm]{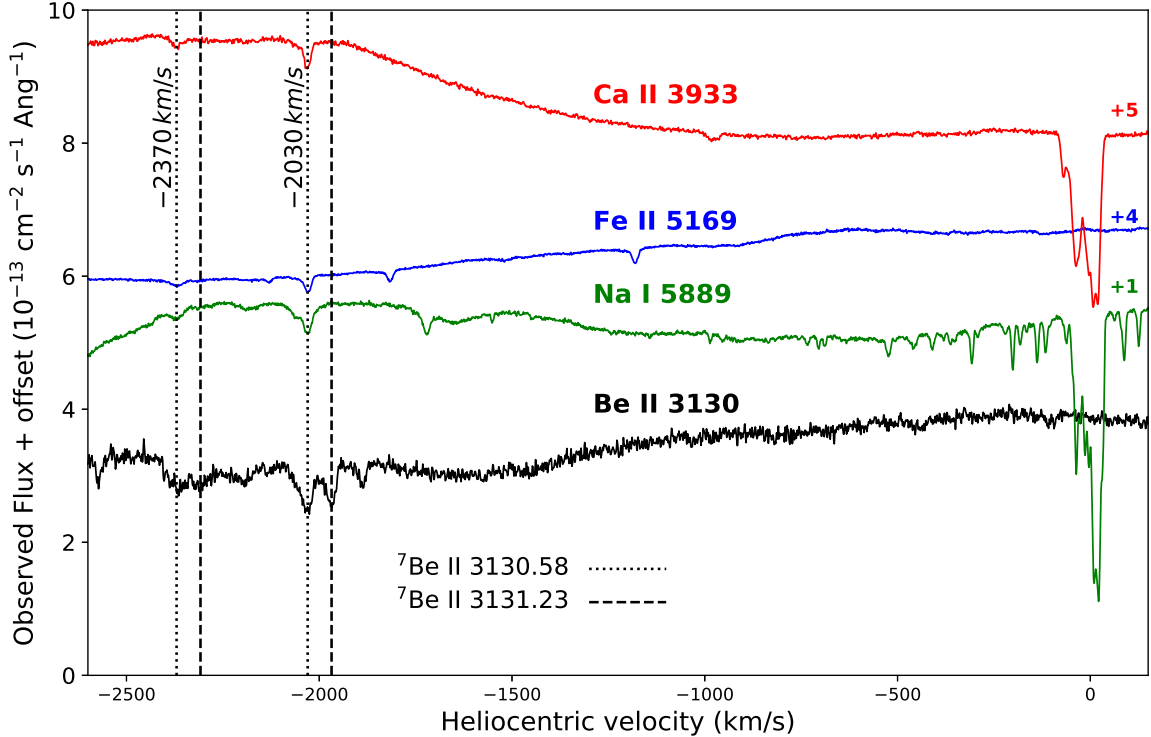}
    \caption{ASASSN-16kt spectrum at Day 8. The figure displays the spectrum around \ion{Be}{II} $\lambda$3130 (black line), \ion{Na}{I} $\lambda$5890 (green line), \ion{Fe}{II} $\lambda$5169 (blue line) and \ion{Ca}{II} $\lambda$3933 (red line) plotted on the velocity scale. The fluxes have been scaled to show the common absorption component at $v_{\rm{exp,1}} = -2030$ km s$^{-1}$, which is shown as vertical dashed (for the \ion{Be}{II} $\lambda$3130 doublet component) and dotted (for the \ion{Be}{II} $\lambda$3131) lines, and at $v_{\rm{exp,2}} = -2370$ km s$^{-1}$, shown as dash-dot lines.}
    \label{fig:5}
\end{figure*}

In addition to the above mentioned low-ionization transitions, we  identified $^7$\ion{Be}{II} $\lambda\lambda$3130,3131, which is blue-shifted at the same velocity as the \ion{Na}{I} D line. The region is relatively free from absorptions of low ionization species and the two lines of the $^7$\ion{Be}{II} are clearly visible. In addition, the P-Cygni profile extends up to velocities $\sim -2370$ km s$^{-1}$, where additional blue-shifted velocities are also observed for \ion{Na}{I} D, \ion{Fe}{II} $\lambda$5168 and \ion{Ca}{II} $\lambda$3933, see Fig. \ref{fig:5}. Following the same approach used for V1369 Cen \citep{Izzo2015} and V5668 Sgr \citep{Molaro2016}, we can estimate the mass abundance of $^7$\ion{Be}{II} ejected in ASASSN-16kt through comparison with the resonance lines of \ion{Ca}{II}. First, we note the absence of the \ion{Ca}{I} $\lambda$4226 resonance line in the Day 8 spectrum, hence we conclude that the Ca in the nova ejecta is largely ionised. Since the resonance transitions differ by 0.80 eV, we can calculate the mass abundance as follows \citep{SpitzerBook}:

\begin{equation}\label{eq:no1}
\frac{A_m(\textrm{Be})}{A_m(\textrm{Ca})} = \Bigg(\frac{EW_{\ion{Be}{II}}}{3130^2} \Bigg/ \frac{EW_{\ion{Ca}{II}}}{3933^2}\Bigg)
 \times \frac{gf_{\ion{Ca}{II}}}{gf_{\ion{Be}{II}}} \times \frac{u_{\textrm{Be}}}{u_{\textrm{Ca}}}
\end{equation}

where $u$ is the atomic mass (40 for Ca and 7 for $^7$\ion{Be}{II}). The oscillator strengths $gf$, are $gf_{\textrm{Be}} = 0.68 + 0.34 = 1.02$ \citep{Biemont1977} and $gf_{\textrm{Ca}} = 1.36$ \citep{Black1972}. We then measured the EWs of the blue-shifted absorption lines associated with both elements. The total contribution due to \ion{Ca}{II} is provided by $EW_{\ion{Ca}{II}} = EW_{\ion{Ca}{II} \lambda3933}$. From the observed spectra, and considering the blue-shifted absorptions at $v_{ej} = -2030,-2370$ km s$^{-1}$ for both transitions, we obtain total EWs of $EW_{\ion{Be}{II}} = 434.3$ m\AA, and $EW_{\ion{Ca}{II}} = 41.4$ m\AA. Considering the formula in Eq. \ref{eq:no1}, we obtain a mass abundance ratio of $A_m(\textrm{Be})/A_m(\textrm{Ca})$ = 3.86, and considering that after 8 days from the initial outburst $\sim 10\%$ of $^7$\ion{Be}{II} has already decayed into $^7$\ion{Li}{I}, we get an initial $^7$\ion{Be}{II} mass abundance produced during the TNR of $A_{m,in}(\textrm{Be})/A_m(\textrm{Ca})$ = 4.25 .

\section{Physical properties from late epochs}

\begin{figure*}
	\includegraphics[width=18cm]{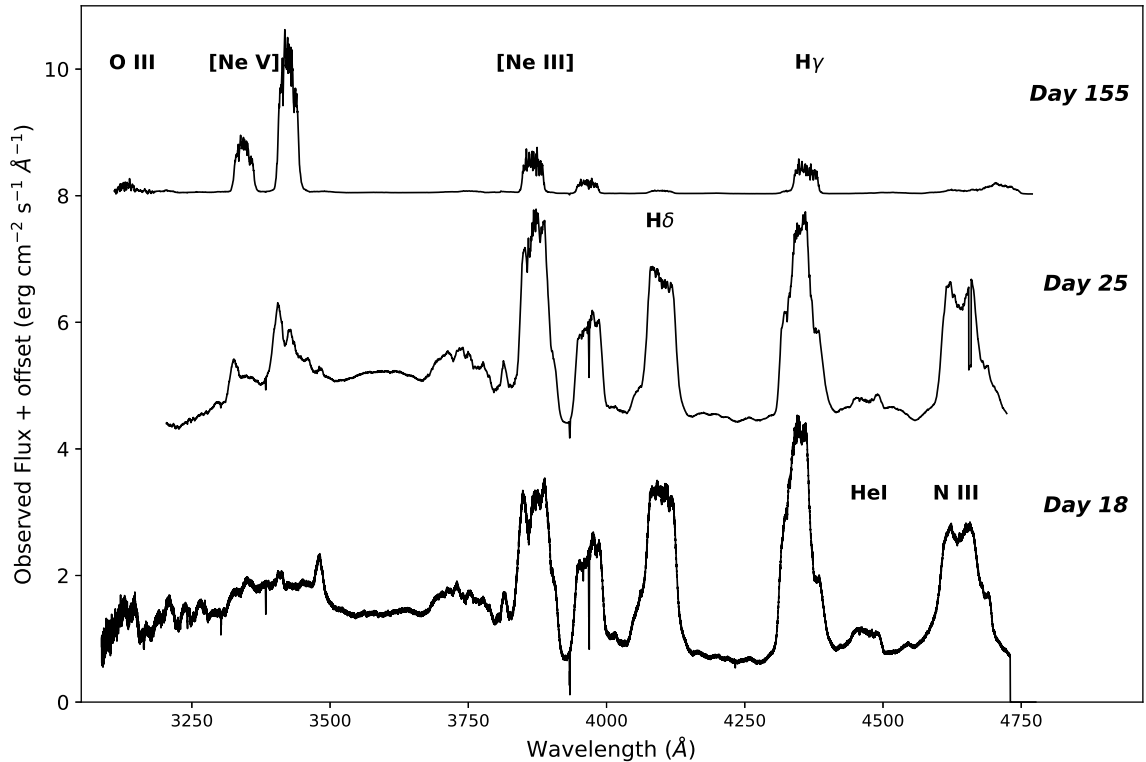}
    \caption{The optical spectra of ASASSN-16kt observed on Day 18 with UVES and on Day 25 and Day 155 with X-Shooter. The flux evolution in time of neon lines as well as of Bowen fluorescence lines as \ion{O}{III} $\lambda$3132 and the blend \ion{N}{III} 4640 is clearly visible. Neon lines are blended with \ion{O}{III} and [\ion{Ne}{III}] lines in the Day 18 and Day 25 spectra.}
    \label{fig:6}
\end{figure*}

All low-ionization absorption features were absent in the UVES spectrum obtained 18 days after the nova's discovery. This spectrum shows the clear presence of \ion{He}{I}  $\lambda\lambda$4471,5876,7065,10830 lines as well as emerging nebular transitions such as [\ion{O}{III}] $\lambda\lambda$4959, 5007 (the auroral [\ion{O}{III}] $\lambda$4363 line is blended with H$\gamma$), and high-ionization iron lines of [\ion{Fe}{VI}] $\lambda\lambda$5176,5677. We also note a faint Bowen fluorescence line of \ion{O}{III} $\lambda$3132 in the spectrum from Day 25, while its detection in the Day 155 spectrum is doubtful. Day 25 spectrum shows also typical forbidden lines of Ne such as [\ion{Ne}{III}] $\lambda\lambda$3869,3967 and [\ion{Ne}{V}] $\lambda\lambda$3346,3426, the latter ones being fainter than [\ion{Ne}{III}] and blended with \ion{O}{III} and [\ion{Ne}{III}] lines. The situation does not change on the Day 25 spectrum, while the spectrum obtained on Day 155 shows a further increase of excitation, marked by the flux increase of [\ion{Ne}{v}] lines, making it even brighter than the H$\alpha$ line, see Fig. \ref{fig:6}, as well as the presence of [\ion{Fe}{VII}] $\lambda$6087 . 

\begin{figure}
	\includegraphics[width=\columnwidth]{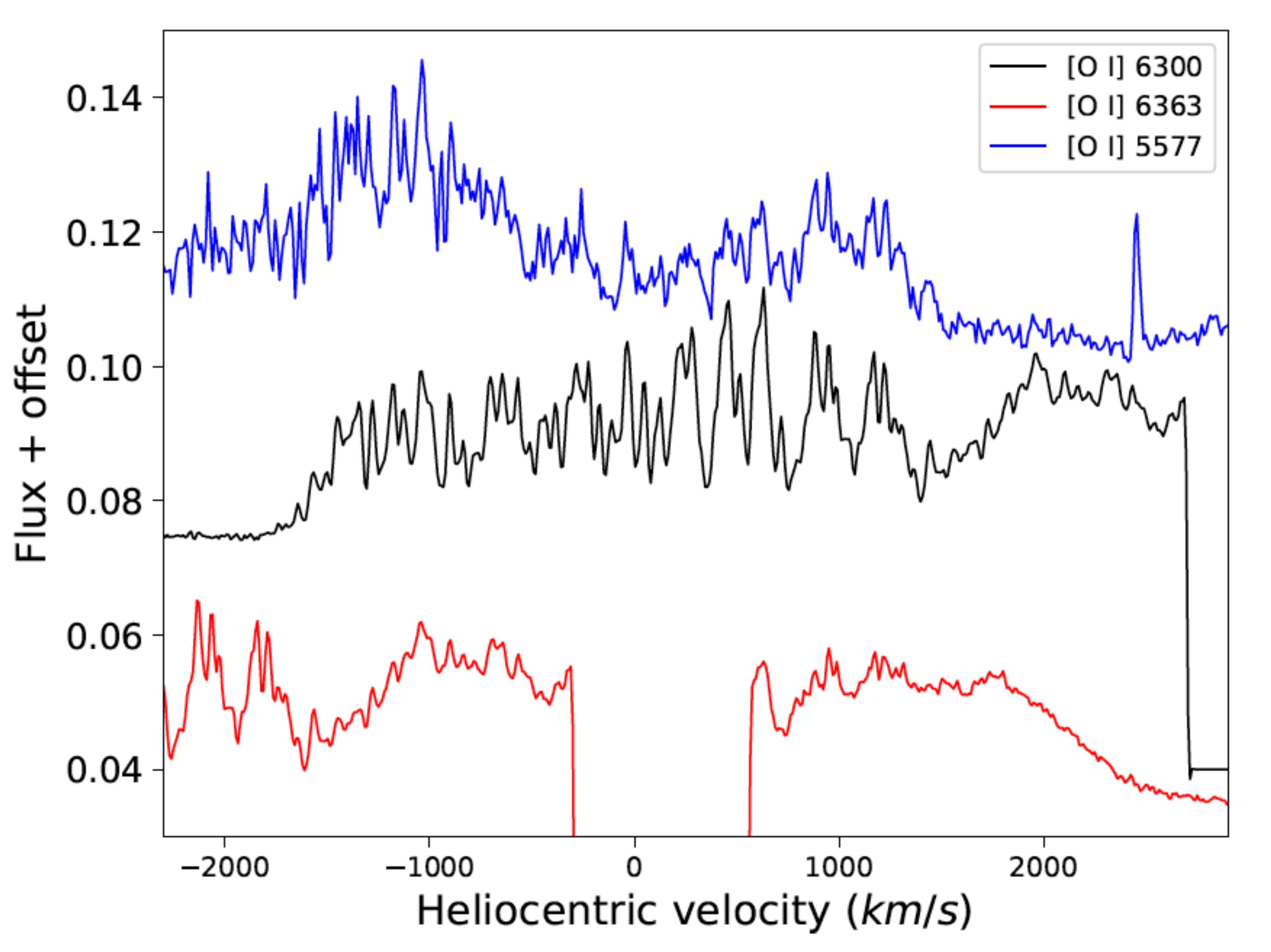}
    \caption{ASASSN-16kt spectrum at Day 155. The figure displays the spectrum centered at \ion{O}{I} $\lambda$6363 (blue line), \ion{O}{I} $\lambda$6300 (black line), and \ion{O}{I} $\lambda$5577 (red line) plotted on the velocity scale. The fluxes have been scaled to show the common structures at the top of their emission line profile, as well as the wider red wing of the [\ion{O}{I}] $\lambda$6363 line, suggesting the presence of an underlying [\ion{Fe}{x}] $\lambda$6375 line. The gap between 6355$\AA$ and 6375$\AA$ is due to the instrument (X-Shooter in this case).}
    \label{fig:7}
\end{figure}

We also detect highly-ionised [\ion{Fe}{x}] $\lambda$6375 blended with [\ion{O}{I}] $\lambda$6363. The intensity of this line is usually 1/3 of the emission of the nearby [\ion{O}{I}] $\lambda$6300, but both lines show a similar observed flux. \citet{Williams1994} explained by existence of cool, high-density blobs of neutral material embedded in the nova shell. In Fig. \ref{fig:7} we show a comparison between the [\ion{O}{I}] $\lambda\lambda$5577,6300 and 6363 lines. While the $\lambda$5577 and $\lambda$6300 lines show similar spiky structures at same ejecta velocities on the top of their emission line profiles, in the [\ion{O}{I}] $\lambda$6363 line the spiky structures are less pronounced with the red wing of the line extending up to larger velocities. This strongly suggests the presence of an additional emission component, which we associate with the [\ion{Fe}{x}] $\lambda$6375 transition. This line is also a good proxy for the presence of a high-energy X-ray photon field, likely originating from a very hot underlying WD \citep{Schwarz2011}. In the Neil Gehrels Swift observatory XRT observations, the maximum of the X-ray supersoft (SSS) emission is around Day 150 from the discovery of the nova \citep{ATEL10632}. One year after the nova discovery the SSS emission is still active but with a count-rate flux in X-rays less than more of one order of magnitude \citep{ATEL10722}.

\begin{figure}
	\includegraphics[width=1.0\columnwidth]{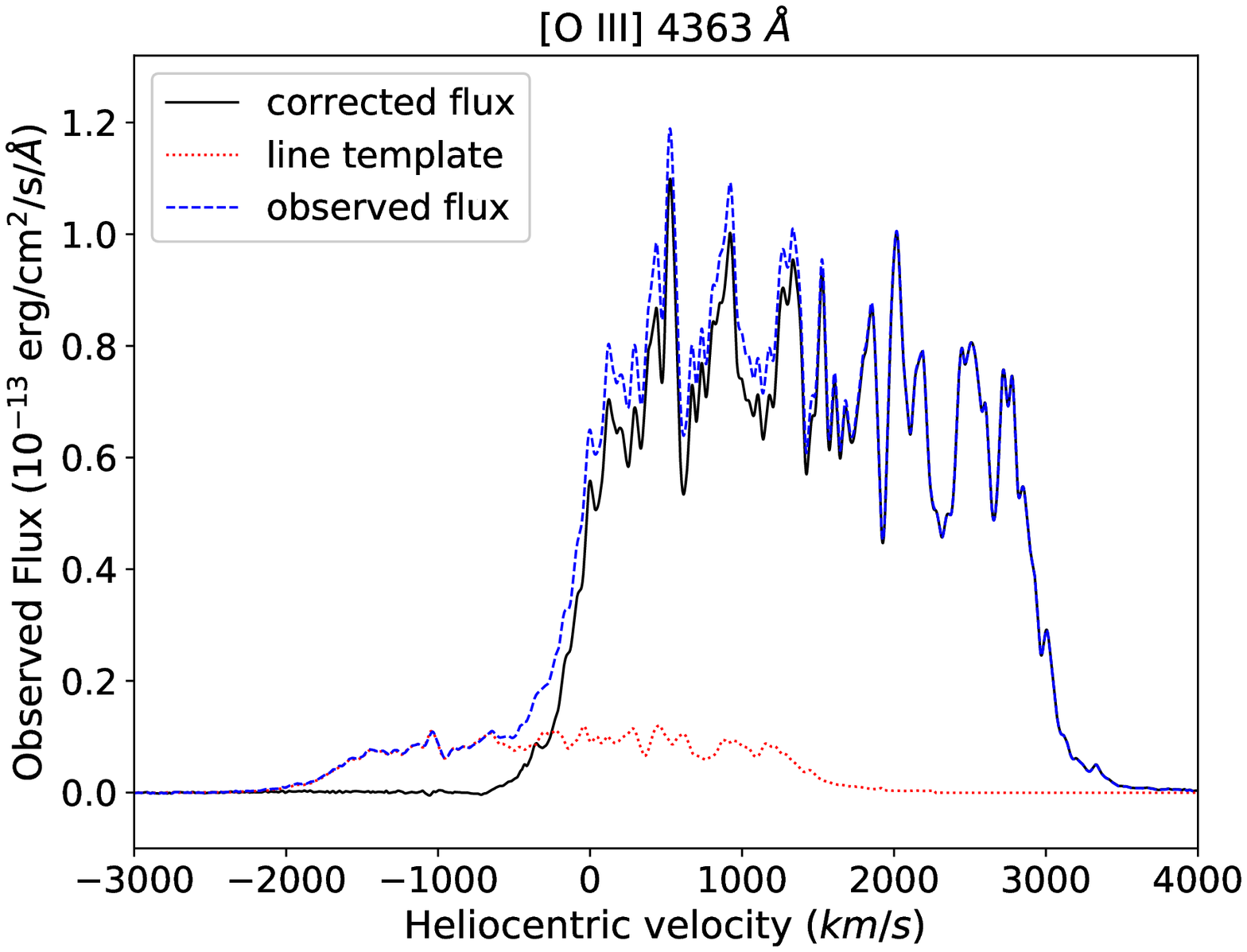}\\
    \includegraphics[width=1.0\columnwidth]{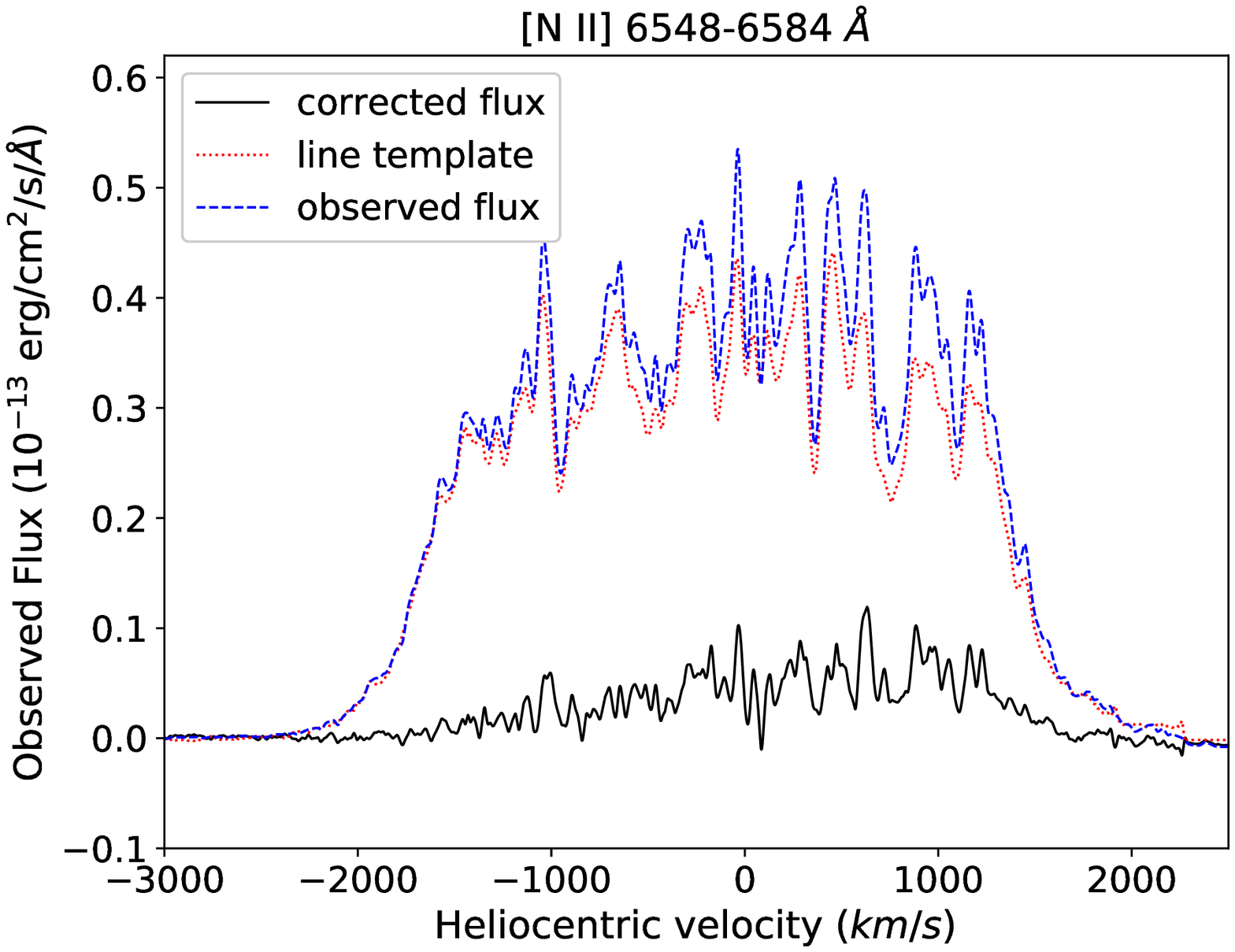}
    \caption{The results of the method used to de-blend [\ion{O}{III}] $\lambda$4363 (upper panel) and the [\ion{N}{II}] $\lambda\lambda$6548,6584 (lower panel) from H$\gamma$ and H$\alpha$ emission line, respectively, as described in Section 5.}
    \label{fig:8}
\end{figure}

The Day 155 spectrum is the last high-resolution spectrum in our dataset, and we can estimate some important physical properties of the nova ejecta from it. We used emission line diagnostics that are usually developed for the analysis of \ion{H}{II} regions as well as planetary nebulae and active galactic nuclei to determine the electron density and the hydrogen mass ejected in the nova outburst. The lack of typical electron density indicators in the spectrum led us to use the nebular lines of [\ion{N}{II}] $\lambda\lambda$ 5755,6548,6584 and [\ion{O}{III}] $\lambda\lambda$ 4363,4960,5007 to determine the electron density $n_e$ and temperature $T_e$. However, the [\ion{N}{II}] $\lambda\lambda$ 6548,6584 and [\ion{O}{III}] $\lambda$4363 lines are blended with H$\alpha$ and H$\gamma$, respectively. To disentangle these lines, we used the H$\beta$ line as a template to rescale their fluxes by taking their blue wing as reference, see Figs. \ref{fig:8}. The results are good for the [\ion{O}{III}] line while the [\ion{N}{II}] lines are not clearly resolved; however for the estimate of the electron density and temperature we need their sum, hence we measured the integrated flux of the observed excess as due to both [\ion{N}{II}] lines. De-reddened fluxes and FWHM for the main emission lines observed in the Day 155 spectrum are reported in Table \ref{tab:3}.

Using the formulas provided in \citet{Osterbrock1989}:
\begin{equation}
\frac{I(\lambda4959) + I(\lambda5007)}{I(\lambda4363)} = \frac{7.90 \exp{32900/T_e}}{1 + 4.5 \times 10^{-4} \frac{n_e}{T_e^{0.5}}},
\end{equation}

\begin{equation}
\frac{I(\lambda6548) + I(\lambda6584)}{I(\lambda5755)} = \frac{8.23 \exp{25000/T_e}}{1 + 4.4 \times 10^{-3} \frac{n_e}{T_e^{0.5}}},
\end{equation}

which are typically used for \ion{H}{II} regions and planetary nebulae but equally applied to the case of novae \citep[see, e.g., ][]{DellaValle2002}, we obtain an electron density of $n_e \sim 2 \times 10^6$ cm$^{-3}$ and a temperature of $T_e \sim 12000$ K.

The presence of very bright [\ion{Ne}{V}] lines suggests a large abundance of neon in the nova ejecta, pointing to a ONe WD progenitor for ASASSN-16kt. We have computed the ionic abundances for Ne using the observed forbidden lines at [\ion{Ne}{III}]$ \lambda\lambda$3869,3967 and [\ion{Ne}{V}]$ \lambda\lambda$3346,3426, respectively. We also note that [\ion{Ne}{IV}] lines are likely blended with other lines, including the \ion{He}{I}$ \lambda$4713, so we have not considered these lines in our computation. In this light, the final results will provide lower limit to the actual ionic abundance of Ne. 

We used the \texttt{NEBULAR.IONIC} task \citep{Shaw1995} within the \texttt{IRAF} package to estimate Ne ionic abundances based on the Hbeta emissivity \citep{DeRobertis1987}. Using the values for the temperature and density obtained before, we obtain the ionic abundances, with respect to hydrogen, of $N([\ion{Ne}{V}])/N(H) = 1.22 \times 10^{-3}$ and $N([\ion{Ne}{III}])/N(H) = 3.59 \times 10^{-4}$. This results in a Ne abundance of $N(Ne) \geq 9.2$, relative to hydrogen, $N(H) = 12$. The solar Ne abundance is $N(Ne) = 8.05 \pm 0.10$ \citep{Lodders2009}. Consequently, we estimate a Ne abundance in the ejecta of ASASSN-16kt which is $\sim$ 14 times the solar one. This value is comparable to that of the ONe nova V1974 Cyg of $N(Ne) \sim 12$ \citep{Gehrz1994}, where the authors used infrared emission lines of Ne to estimate the abundance.

\subsection{The ejected mass of $^7$Be II}

Following \citet{Mustel1970} we can estimate the mass of the hydrogen envelope ejected in the outburst. To this aim, we first need the knowledge of the geometry of the nova ejecta. It is well known that nova ejecta are not spherical \citep{Cohen1983,Slavin1995,Gill1998,Gill2000,Sahman2015}, an evidence confirmed also at radio frequencies \citep{Chomiuk2014}. In fact, nova ejecta frequently show equatorial rings and/or polar caps condensations, which make them more similar to oblate or prolate spheroids than spheres. In what follows, we assume a spherical geometry as a first-order approximation for the nova ejecta, but the other driving parameter is given by the filling factor, which quantifies the degree of condensations inside the nova ejecta. Consequently, we first assume that the volume $V$ of the ejecta can be approximated by a spherical shell with radius $R = v_{ej} \Delta t$, where $v_{ej} \eqsim 2300$ km s$^{-1}$ is the expansion velocity measured from the Day 155 spectrum and $\Delta t = 13.4$ Ms is the time since outburst, and with thickness of the shell $\delta = a R \epsilon$, where $a = 0.17$ corresponds to the observed ratio between the width of Balmer lines P-Cygni profiles and their blue-shifted velocity, while $\epsilon$ is the filling factor. The observed flux of the recombination Balmer lines (in this case we consider H$\beta$) is then given by:
\begin{equation}
I_{H\beta} = \Big[\Big(\frac{4 \pi j_{H\beta}}{n_e n_p}\Big) n_e n_p V \Big] \frac{1}{4\pi d^2},
\end{equation}
with the term $\frac{4 \pi j_{H\beta}}{n_e n_p}$ being the H$\beta$ emissivity \citep{Osterbrock1989} and $d$ the distance to the nova. Assuming $n_e = n_p$, we can estimate the filling factor from the previous formula , being $\epsilon = 0.4$ at this epoch, and then the mass of the hydrogen ejected in the outburst $M_{H,ej} \approx 5.5 \times 10^{28}$ g, corresponding to $M_{H,ej} = 2.7 \times 10^{-5} M_{\odot}$.

We can finally compute the mass of Ca following the semi-empirical analysis developed by \citet{Izzo2015}, where different nova composition models have been considered to estimate the mass of synthesized elements compared to the ejected H mass. For a generic ONe WD progenitor with a mass varying from 1.0 and 1.35 M$_{\odot}$, the mass of Ca varies between $(1.66 - 1.82) \times 10^{-5}$ of the total ejecta mass, and between $(5.1 - 6.8) \times 10^{-5}$ the ejected H mass, according to the numerical results of \citet{Politano1995}. These values correspond to an ejected Ca mass of $(1.38 - 1.83) \times 10^{-9}$ M$_{\odot}$. Assuming the ejected H mass computed above and a mass abundance ratio of 4.24 for $^7$\ion{Be}{II} with respect to Ca as estimated in the previous section, we obtain a total mass of $^7$\ion{Be}{II} ejected in ASASSN-16kt of $(5.9 - 7.7) \times 10^{-9} M_{\odot}$, which is of the same order of magnitude of the $^7$\ion{Be}{II} mass estimated in V5668 Sgr \citep{Molaro2016} and one order of magnitude larger than the Li mass estimated in V1369 Cen \citep{Izzo2015}, but in this latter case only the observations of the $^7$Li $\lambda$6708 line were considered.

\begin{table}
\centering
\caption{De-reddened fluxes and FWHM for bright emission lines measured in the UVES spectrum of Day 155. Fluxes are in units of 10$^{-13}$ erg cm$^{-2}$ s$^{-1}$ while FWHM are in km s$^{-1}$.}
\label{tab:3}
\begin{tabular}{l c c c}     
\hline
Line & $\lambda_0$ & Flux & FWHM \\
\hline
$[\ion{Ne}{V}]$ & 3346 &  59.40 & 2600\\
$[\ion{Ne}{V}]$ & 3426 &  173.33 & 2600\\
$[\ion{Ne}{III}]$ & 3869 &  42.124 & 2550\\
$[\ion{Ne}{III}]$ & 3967 &  13.67 & 2600\\
$[\ion{O}{III}]$ & 4363 &  37.60 & 3100\\
H$\beta$ & 4861 & 7.43 & 2550\\
$[\ion{O}{III}]$ & 4959 & 18.27 & --\\
$[\ion{O}{III}]$ & 5007 & 58.75 & 3100\\
$[\ion{N}{II}]$ & 5755 & 3.58 & 2550\\
\ion{He}{I} & 5876 & 0.51 & 3000\\
$[\ion{N}{II}]$ & 6548+6584 & 3.06 & --\\
H$\alpha$ & 6562 & 22.24 & 2550\\
\ion{He}{I} & 10830 & 12.29 & 2500\\
\ion{He}{I} & 12528 & 1.29 & 3700\\
Pa$\beta$ & 12818 & 1.73 & 2800\\
\hline
\end{tabular}
\end{table}

\section{Discussions} 

All novae with positively identified $^7$\ion{Be}{II} and/or Li  to date have been CO novae. Progenitors of CO novae have generally smaller masses compared to oxygen-neon (ONe) novae \citep{IbenTutukov1985,Straniero2016}. Moreover, a less massive WD progenitor needs a larger amount of accreted matter to reach the critical pressure to hence ignite the TNR \citep{Fujimoto1982,Prialnik1982,Truran1986}. Consequently, CO novae are characterized by more massive ejecta than ONe novae (typically 10 times larger or more). In turn, the mass of $^7$\ion{Be}{II} produced in ONe novae should be less than that produced in CO novae, as expected from numerical simulations \citep{Hernanz1996}. The $^7$Be mass estimated for ASASSN-16kt is indeed less than the $^7$Be mass in V5668 Sgr \citep{Molaro2016}, which is in agreement with the above picture. 

Another consequence of the above scenario is that the luminosity a nova reaches at maximum correlates with the WD progenitor mass: the larger the WD mass, the smaller the accreted envelope mass, as well as the total binding energy \citep{Truran1986}. Moreover, for massive WDs the luminosity also anti-correlates with the accretion rate: the newly-discovered class of very fast and faint novae \citep{Kasliwal2011} represent the main outcome of this physical scenario for CNe.

An analytical formulation relating the absolute $B$-band magnitude at maximum and the WD progenitor mass was proposed by \citet{Livio1992}: for a distance of $d = 10.0$ kpc, we have $M_B = -9.1$ mag, after correcting the observed value $B = 6.86$ mag for reddening, and then a WD mass of $M_{WD} = (1.20 \pm 0.05)$ M$_{\odot}$. Assuming this value for the WD progenitor mass, according to \citet{Truran1986} we obtain a recurrence time for ASASSN-16kt of $T_{\textrm{rec}} \sim 3 \times 10^4$ years. When applying the same formulae to V5668 Sgr, we find a WD progenitor mass of 0.84 M$_{\odot}$, and a recurrence time of $T_{\textrm{rec}} \sim 3 \times 10^5$ years, which is almost one order of magnitude larger than ASASSN-16kt. If we take into account the estimated $^7$Be yield produced by ASASSN-16kt and its recurrence time, and the observed evidence that fast novae represent $\sim 30\%$ of the total novae in the Galaxy, we conclude that ONe novae produce a similar quantity of $^7$\ion{Be}{II} than the one produced by CO novae, like V5668 Sgr.

\section{Conclusions}

ASASSN-16kt is a very fast nova ($t_2 = 9.9 \pm 0.1$ days) at a distance of $d = 10.0 \pm 1.5$ kpc. The first spectra obtained five to eight days from its discovery reveal that ASASSN-16kt is an hybrid nova in the CTIO spectral classification \citep{Williams1992}: the identification of \ion{Fe}{II} lines is due to the presence of blue-shifted absorption lines at the same expanding velocities reported for other lines, like \ion{Na}{I}, \ion{Ca}{II} and also $^7$\ion{Be}{II}. The two lines of the $^7$\ion{Be}{II} doublet are clearly seen and not contaminated by blends for the components at $v_{ej,1} = -2030$ km/s and $v_{ej,2} = -2370$ km/s. Moreover, the late spectra show it is a ONe nova, due to the presence of very bright [\ion{Ne}{III}] and  [\ion{Ne}{V}] lines and from the Ne abundance value derived from their analysis. Given its height above the galactic plane ($h = 1.7$ kpc), it is also a bulge nova, according to \citet{DellaValle1998}. However, bulge novae are characterized by distinct properties, such as being slower and fainter than ASASSN-16kt, and are usually characterized by a typical \ion{Fe}{II} spectral class. ASASSN-16kt in this sense would represent an interesting outlier of this sub-class.

The detection of $^7$Be, which decays completely into $^7$Li through electron capture \citep{Giraud2007}, in the early spectra of the ONe nova ASASSN-16kt further confirms that novae represent the main Li-factories in the Galaxy. The estimate of the mass of $^7$\ion{Be}{II} ejected is in line with other values published in literature \citep{Izzo2015,Molaro2016}, implying that the contribution of ONe novae to Galactic lithium enrichment is not negligible compared to CO novae, but given also their higher recurrence frequencies, ONe novae produce a similar quantity of $^7$\ion{Be}{II} as CO novae. The evidence that $^7$\ion{Be}{II} has been identified only at early epochs in ASASSN-16kt implies that the detection of $^7$\ion{Be}{II}, is possible only during the epochs of the nova characterized by the presence of low-ionization elements, i.e. during the maximum peak and the diffuse-enhanced phase \citep{McLaughlin1942,Payne1957}, or following the CTIO classification \citep{Williams1992} when \ion{Fe}{II} lines are present in the spectrum. 

A more precise characterization of the $^7$\ion{Be}{II} and Li mass produced during a nova outburst as a function of the progenitor WD mass is still needed, in order to precisely quantify the total Galactic yield of lithium provided by novae. The large quantity of $^7$\ion{Be}{II} measured in the few novae observed so far suggests a very large overabundance if we assume that all novae produce approximately the same quantity of $^7$Be/$^7$Li in each outburst. These results represent an important input for numerical simulations of TNR nucleosynthesis: according to the results presented in \citet{Jose1998}, the quantity of $^7$\ion{Be}{II}, and hence Li, produced during the TNR varies in between $\sim 10^{-5}-10^{-6}$ times the ejected H mass in the outburst, which is a value $\sim$ an order of magnitude lower than the observed quantity for both CO and ONe novae.  

These findings imply that classical novae produced a huge quantity of Li, more than the value of 150 $M_{\odot}$ of Li estimated to be in the Milky Way \citep{Fields2014}. If confirmed by more detections of Be and Li in future classical novae, we could even be faced with a new ''lithium problem'': since $^7$Be decays only via electron-capture producing $^7$Li \citep{Giraud2007}, some depletion mechanisms must work in the immediate phases after its formation in order to destroy a sufficient quantity of $^7$Li, without claiming the existence of more exotic physics (but see \citealp{Krasznahorkay2016}). Additional detections of $^7$\ion{Be}{II} and/or $^7$Li combined with a detailed study of the ejecta properties, like its degree of asphericity \citep{Ribeiro2011,Chesneau2012,Shore2013}, are needed to quantify the nova Galactic yield, and then to infer the presence of mechanisms acting during the TNR, or in the nova ejecta, that can deplete freshly-formed Li, and finally explain the over-abundances of Li observed in young stellar populations \citep{Spite1990}.

\section*{Acknowledgements}

We thank the referee for her/his very constructive comments that have largely improved the paper. We thank Andreas Kaufer for his prompt reply to our ESO-DDT request and recognize the precious support of John Pritchard for the success of the observations. We also thank Nestor Espinosa for his contribution to the success of PUCHEROS observations. We acknowledge with thanks the variable star observations from the AAVSO International Database contributed by observers worldwide and used in this research. LI, CT, ZC and AdUP acknowledge support from the Spanish research project AYA 2014-58381-P. CT and AdUP furthermore acknowledge
support from Ramón y Cajal fellowships RyC-2012-09984 and RyC-
2012-09975. ZC also acknowledges support from the Juan de la Cierva Incorporaci\'on fellowship IJCI-2014-21669. LV acknowledge support from CONICYT through project Fondecyt n. 1171364. AZ acknowledge support from Conicyt through the "Beca Doctorado Nacional" 21170536. Support for JLP is provided in part by FONDECYT through the grant 1151445 and by the Ministry of Economy, Development, and Tourism’s Millennium Science Initiative through grant IC120009, awarded to The Millennium Institute of Astrophysics, MAS.





\bibliographystyle{mnras}


\bsp	
\label{lastpage}
\end{document}